\documentclass[numreferences]{kluwer}

\newdisplay{guess}{Conjecture}

\begin{document}

\begin{article}

\begin{opening}

\title{Extended Discrete KP Hierarhy and its Reductions from a Geometric
Viewpoint\thanks{This
work was supported in part by INTAS grant 2000-15.}}

\author{ANDREI K. \surname{SVININ}}

\runningauthor{ANDREI K. SVININ}

\runningtitle{Extended discrete KP hierarhy}

\institute{Institute of System Dynamics and Control Theory, Siberian
Branch of Russian Academy of Sciences, P.O. Box 1233, 664033 Irkutsk,
Russia}

\date{}

\begin{abstract}
We interpret the recently suggested extended discrete KP (Toda lattice)
hierarchy from a geometrical point of view. We show that
the latter
corresponds to the union of invariant submanifolds $S_0^n$ of the system
which is a chain of infinitely many copies of Darboux-KP hierarchy, 
while the intersections $S_0^n\cap S_{l-1}^{ln-r}$
yield a number of reductions to $l$-field lattices.
\end{abstract}


\end{opening}

\section{Introduction}

In our previous work \cite{svinin1} we introduced a so-called extended
discrete KP (edKP) hierarchy. Let us give a sketch of the picture
of this within the framework of a Sato-like approach.

One considers the phase-space ${\cal M}$
which consists of the entries of
diagonal matrices $w_k = (w_k(i))_{i\in{\bf Z}},\;k\ge 1$. To each point
of ${\cal M}$, one associates an infinite set of ``wave" operators
\[
\{S^{(n)} = I + \sum_{k\ge 1}w_kz^{k(n-1)}\Lambda^{-kn}:\; n\ge 1\}
\]
where $\Lambda$ is a shift operator acting as $(\Lambda f)(i) = f(i+1)$.
The flow on ${\cal M}$ with respect to the parameter $t_p^{(n)}$ is defined
by the equation of motion\footnote{We use here the shorthand notation
$S$ for $S^{(n)}$ and $Q$ for $Q^{(n)}$.}
\begin{equation}
z^{(n-1)p}\frac{\partial S}{\partial t^{(n)}_p}
= Q_{+}^{pn}S-S\Lambda^{pn}
\label{eq-m}
\end{equation}
with  the Lax operator $Q\equiv S\Lambda S^{-1}$.
Evolutions of $S$ yield
evolutions of $Q$ in the form of the Lax equations
\begin{equation}
z^{(n-1)p}\frac{\partial Q}{\partial t^{(n)}_p}
= [Q_{+}^{pn}, Q].
\label{lax-eq-m}
\end{equation}

The collection of evolution parameters $\{t_p^{(n)},\; n\ge 1,\; p\ge 1\}$
can be grouped  into infinite sets $t^{(n)}\equiv (t_1^{(n)}\equiv x^{(n)},
t_2^{(n)},...)$ --- multi-times. Any subsystem of edKP hierarchy (\ref{eq-m})
corresponding to fixed $n$  is reffered to, due to  \cite{svinin2} as $n$th
discrete KP hierarchy. In particular, in the case $n=1$, (\ref{eq-m})
(or (\ref{lax-eq-m})) gives equations of ordinary discrete KP (Toda lattice)
hierarchy.
It was shown in \cite{svinin1}, \cite{svinin2} that $n$th discrete KP in fact
is equivalent to bi-infinite sequence of continuous KP hierarchy copies
with corresponding
$\partial^{(n)}$-Lax operators ${\cal Q}_i$ glued together by compatible
gauge transformations. From this we learn that the flows of $n$th discrete
KP hierarchy commute with each other.

The aim of this Letter is to discuss edKP hierarchy from geometric point of
view which was proposed in a series of the papers \cite{cfmp}.
The main inspiration of this work is the paper \cite{mpz}, where the authors
exploiting the concept of Darboux covering (see also \cite{1,2}) unify many
notions commonly used
in KP theory. Among these are Miura and Darboux transformations, Krichever
rational reductions etc.. In particular they showed that well known discrete
KP hierarchy can be interpreted as a restriction of Darboux--KP (DKP) chain
(\ref{DKPi}) on some invariant submanifold.

In the present paper we generalize this result by showing more general class
of invariant submanifolds of DKP chain phase-space. This allows us to give
simple geometric interpretation of edKP hierarchy and its reductions.
We show that multi-times $t^{(1)}, t^{(2)},...$, in fact, corresponds to
different invariant submanifolds $S_0^1, S_0^2,...$ We believe that geometrical
point of view allows offer clearer interpretation of the role of different
multi-times in definition of edKP hierarchy (\ref{eq-m}).

The Letter is organized as follows. In Section 2, we prove theorem which
provides us by a class of invariant submanifolds of DKP chain $S_l^n$.
In Section 3, we show that submanifold $S^n_0$ corresponds to $n$th discrete
KP. Finally, Section 4 is concerned with a class of reductions of edKP
which are shown to be tantamount to intersections of some invariant
submanifolds.

\section{DKP chain}

To begin with, let us recall the notion of the KP hierarchy following along the
lines suggested
in \cite{cfmp}. The main idea behind this approach
is in fact an outcome
of bihamiltonian formalism of evolution equations defined on special
infinite-dimensional Poisson manifold ${\cal P}$. The points of ${\cal P}$
are parametrized by infinite sets of smooth scalar functions
$\{h_k(x),\; {k\ge 1}\}$ of spatial variable $x\in\dR$. To every point of
${\cal P}$ one
associates monic Laurent series
\[
h(z, x) = z + \sum_{k\ge 1}h_k(x)z^{-k}
\]
with $z$ being considered as the point of unit circle
and Fa\`a di Bruno iterates $h^{(k)}$ constructed, starting from
$h^{(0)}=1$, with the help of recurrence relation $h^{(k+1)} =
\partial h^{(k)}/\partial x + hh^{(k)}$. Then one introduces negative
iterates $h^{(-1)}, h^{(-2)},...$ by solving recursively the equations
\[
\frac{\partial h^{(-1)}}{\partial x} + h h^{(-1)} = 1,
\]
\[
\frac{\partial h^{(-2)}}{\partial x} + h h^{(-2)} = h^{(-1)}
\]
and so on.
Having in mind Fa\`a di Bruno iterates, one defines polarization of
Laurent series space
\[
L = \left\{
\sum_{-\infty\ll k<\infty}l_kz^{-k}
\right\} = H_{+}\oplus H_{-}
\]
with $H_{+} = {\rm span}(h^{(k)} : {k\ge 0})$ and $H_{-} =
{\rm span}(h^{(k)} : {k< 0})$. The currents
$H^{(p)}$ are defined as projection $H^{(p)} = \pi_{+}(z^p)$
with respect to this polarization. More explicitly, one can write
\[
H^{(p)} = h^{(p)}  + \sum_{k=0}^{p-2}r_k^p[h]h^{(k)},
\]
where the coefficients $r_k^p$ can be uniquely specified so that $H^{(p)}$
will be of the form $H^{(p)} = z^p + {\cal O}(z^{-1})$. For example, one has
\[
H^{(1)} = h = z + h_1z^{-1} + {\cal O}(z^{-2}),
\]
\[
H^{(2)} = h^{(2)} - 2h_1 = z^2 + (h_{1x} + 2h_2)z^{-1} + {\cal O}(z^{-2})
\]
and so on.
Then the KP equations are defined, in this
approach, in the form of local conservation laws
\[
\partial_p h = \partial H^{(p)},\; p\ge 1
\]
with $\partial_p\equiv \partial/\partial t_p$ and
$\partial\equiv \partial/\partial x$. \footnote{For any function $a$ of
evolution parameters we will write $\partial a$ and $\partial_p a$
to denote derivatives with respect to corresponding variable.}
The series $h(z)$, in this connection, is treated as generating function of
Hamiltonian densities. It is worth mentioning the invariance property
\cite{cfmp}
\begin{equation}
(\partial_p +  H^{(p)})H_{+}\subset H_{+}
\label{x}
\end{equation}
which will be used in proof of Proposition 2.

Using fundamental concept of Darboux covering, in \cite{mpz},
was introduced the notion of DKP hierarchy
\[
\begin{array}{l}
\partial_p h = \partial H^{(p)} \\
\partial_p a = a(\tilde{H}^{(p)} - H^{(p)})
\end{array}
\]
which turned out to be instrumental, in particular, in constructing
invariant submanifolds of KP hierarchy.
Here $a(z, x) = z + \sum_{k\ge 0}a_k(x)z^{-k}$ and $\tilde{H}^{(p)}$
is the current $H^{(p)}$ evaluated at the point $\tilde{h} = h + a_x/a$.

Fixing any integer $i_0$ one identifies $h=h(i_0)$, $\tilde{h}=h(i_0+1)$ and
$a=a(i_0)$. With  $\tilde{h}$, one defines ``new" DKP hierarchy
\[
\begin{array}{l}
\partial_p \tilde{h} = \partial \tilde{H}^{(p)} \\
\partial_p \tilde{a} = \tilde{a}(\tilde{\tilde{H}}^{(p)} - \tilde{H}^{(p)}).
\end{array}
\]
Again one identifies $\tilde{\tilde{h}} = \tilde{h} + \tilde{a}_x/\tilde{a}
=h(i_0+2)$ with $\tilde{a} = a(i_0+1)$. Continuation of this process gives
infinitely many copies of DKP labeled by $i_0, i_0+1, i_0+2,...$
It is clear this process can be also performed backwards with respect to
$i$. In this way one
arrives at the system
\begin{equation}
\begin{array}{l}
\partial_p h(i) = \partial H^{(p)}(i) \\
\partial_p a(i) = a(i)(H^{(p)}(i+1) - H^{(p)}(i)).
\end{array}
\label{DKPi}
\end{equation}
In what follows, for brevity, we call this system DKP chain.

This differential-difference system was considered in \cite{mpz}.
It was claimed
there that submanifold $S_l$ defined by the constraint
$\pi_{-}(i)\left(z^la(i)\right) = 0$\footnote{Here and in Lemma 1
$\pi_{-}(i)$ stands for projection of any Laurent series on the
space $H_{-}(i)$.}
is invariant with respect to (\ref{DKPi}). In the following we show more
general result stating in the Theorem 1 below, but before let us
prove following technical lemma.

\newtheorem{theoremdemo}{Lemma}
\begin{theoremdemo}
For submanifold $S_l^n$ of DKP lattice phase-space defined
by the constraint
\begin{equation}
\pi_{-}(i)\left(z^{l-n+1}a^{[n])}(i)\right) = 0
\label{constraint}
\end{equation}
the following holds
\begin{eqnarray}
1) & \left(\partial_p + H^{(p)}(i)\right)z^{l-n+1}a^{[n]}(i)\in H_{+}(i), \label{2}\\
2) & z^{p(l-n+1)}a^{[pn]}(i)\in H_{+}(i),\; \forall p\ge 0. \label{3}
\end{eqnarray}
\end{theoremdemo}

Let us pause a bit to give some remarks conserning our notation. In (\ref{constraint})
$a^{[k]}(i)$'s are discrete Fa\`a di Bruno iterates defined by
recurrence relation $a^{[k+1]}(i) = a(i)a^{[k]}(i+1)$ with
$a^{[0]}(i)\equiv 1$. From the definition of $a^{[k]}(i)$'s it follows the obvious
relation
\begin{equation}
a^{[k_1]}(i+k_2)a^{[k_2]}(i) = a^{[k_1]}(i)a^{[k_2]}(i+k_1) =
a^{[k_1+k_2]}(i),\;\; \forall k_1, k_2\in{\bf Z}.
\label{obvrel}
\end{equation}
Observe that from the definition of $a^{[k]}(i)$'s and (\ref{DKPi}) one gets
\begin{equation}
\partial_p a^{[n]}(i) = a^{[n]}(i)(H^{(p)}(i+n) - H^{(p)}(i)),\;
\forall n\in{\bf Z}.
\label{y}
\end{equation}

\begin{pf*}{Proof (of Lemma1)}
First, let us prove by induction that by virtue of DKP chain equations
(\ref{DKPi}), the identity
\begin{equation}
(\partial + h(i))^k
z^{l-n+1}a^{[n]}(i) =
z^{l-n+1}a^{[n]}(i)h^{(k)}(i+n),\;\; \forall k\ge 0
\label{identity}
\end{equation}
holds. For $k=0$ (\ref{identity}) is clear. Suppose that (\ref{identity})
is true for some $k\ge 0$. Then by virtue of (\ref{DKPi}) and (\ref{y})
we have
\[
(\partial + h(i))^{k+1}z^{l-n+1}a^{[n]}(i) =
z^{l-n+1}a^{[n]}(i)(h(i+n) - h(i))h^{(k)}(i+n)
\]
\[
+ z^{l-n+1}a^{[n]}(i)\partial h^{(k)}(i+n) +
z^{l-n+1}a^{[n]}(i)h(i)h^{(k)}(i+n)
\]
\[
= z^{l-n+1}a^{[n]}(i)h^{(k+1)}(i+n).
\]
So (\ref{identity}) is proved. Since, by definition, on $S_l^n$ we have
$z^{l-n+1}a^{[n]}(i)\in H_{+}(i)$, from (\ref{identity}) we get
\begin{equation}
z^{l-n+1}a^{[n]}(i)H_{+}(i+n)\subset H_{+}(i).
\label{identity1}
\end{equation}
Now taking into account (\ref{identity1}) we get
\[
(\partial_p + H^{(p)}(i))z^{l-n+1}a^{[n]}(i) =
z^{l-n+1}a^{[n]}(i)(H^{(p)}(i+n) - H^{(p)}(i))
\]
\[
+ z^{l-n+1}a^{[n]}(i)H^{(p)}(i) =
z^{l-n+1}a^{[n]}(i)H^{(p)}(i+n)\in H_{+}(i).
\]
So we have shown (\ref{2}). Formula (\ref{3}) can be easily proven
by induction using the relation
\[
z^{l-n+1}a^{[n]}(i)\cdot z^{p(l-n+1)}a^{[pn]}(i+n) =
z^{(p+1)(l-n+1)}a^{[(p+1)n]}(i)
\]
which follows from (\ref{obvrel}). Therefore the lemma is proved.
\end{pf*}

Now we are in position to prove

\newtheorem{theoremdemo1}{Theorem}
\begin{theoremdemo1}
The submanifold $S_l^n$ defined by the constraint
(\ref{constraint}) is tangent with respect to the flows of DKP chain
(\ref{DKPi}).
\end{theoremdemo1}

\begin{pf}
In what follows, it is covenient to define the functions
$a_k^{[n]}(i)$ through
the relation
\[
a^{[n]}(i) = z^n + \sum_{k\ge 0}a_k^{[n]}(i)z^{n-k-1}
\]
which are evidently polynomials of $a_k(i)$.
Then
\[
z^{l-n+1}a^{[n]}(i) = z^{l+1} + \sum_{k\ge 0}a_k^{[n]}(i)z^{l-k}.
\]
With these relations the constraint (\ref{constraint}), as can be easily
checked, can be written as
\begin{equation}
z^{l-n+1}a^{[n]}(i) = H^{(l+1)}(i) + \sum_{m=0}^la_m^{[n]}(i)H^{(l-m)}(i).
\label{z}
\end{equation}
So, to prove the theorem one has to show that on $S_l^n$ the identity
\[
\partial_p\left(z^{l-n+1}a^{[n]}(i) -
H^{(l+1)}(i) - \sum_{m=0}^la_m^{[n]}(i)H^{(l-m)}(i)
\right) = 0
\]
holds. The relation (\ref{2}) from the above Lemma, implies
\[
\partial_p\left(z^{l-n+1}a^{[n]}(i)\right) +
H^{(p)}(i)\left(
H^{(l+1)}(i) + \sum_{m=0}^la_m^{[n]}(i)H^{(l-m)}(i)
\right) \in H_{+}(i).
\]
Taking into consideration that $H^{(m)}(i)H^{(p)}(i) + \partial_p H^{(m)}(i)
\in H_{+}(i)$, for all $m, p \ge 1$, which follows from invariance
property (\ref{x}), one can write
\[
\partial_p\left(z^{l-n+1}a^{[n]}(i) - H^{(l+1)}(i)\right) -
\sum_{m=0}^la_m^{[n]}(i)\partial_pH^{(l-m)}(i)\in H_{+}(i).
\]
After a look at positive powers of $z$, one gets
\[
\partial_p\left(z^{l-n+1}a^{[n]}(i) - H^{(l+1)}(i)\right) -
\sum_{m=0}^la_m^{[n]}(i)\partial_pH^{(l-m)}(i)
\]
\[
= \sum_{m=0}^l\left(\partial_pa_m^{[n]}(i)\right)H^{(l-m)}(i).
\]
Therefore the theorem is proved.
\end{pf}

\section{edKP hierarchy from geometric point of view}

Let us discuss restriction of DKP chain on submanifold $S_0^n$ with $n\ge 1$,
which is defined by relation
\begin{equation}
h(i) = z^{1-n}a^{[n]}(i) - a_0^{[n]}(i) =
z^{1-n}a^{[n]}(i) - \sum_{s=1}^na_0(i+s-1)
\label{constraint2}
\end{equation}
which is specification of (\ref{z}).
Observe that restricted DKP lattice equations, thanks to (\ref{constraint2}),
entirely can be rewritten in terms of coordinates $a_k(i)$.
From this point, we need to introduce
multi-times $t^{(1)}, t^{(2)},...$ to differ the motion
on invariant submanifolds $S_0^1, S_0^2,...$.

In what follows, it is useful to introduce, besides $a_k(i)$, coordinates
$q_k^{(n,r)}(i)$, for any integer $r$, by the relation\footnote{Note that
(\ref{relation}) is simply the relation which serves as the map
$\{a_k(i)\}\rightarrow\{q_k^{(n,r)}(i)\}$ without reference to any invariant
submanifold.}
\begin{equation}
z^r = a^{[r]}(i) + \sum_{k\ge 1}q_{k-1}^{(n,r)}(i)z^{k(n-1)}a^{[r-kn]}(i).
\label{relation}
\end{equation}
Since $z^{k(n-1)}a^{[r-kn]}(i)$ is a monic Laurent series of degree $(r-k)$,
the coefficients $q_k^{(n,r)}(i)$ are properly defined as polynomial functions
of coordinates $a_k(i)$.

Let us define attached to any $n\ge 1$ the set of currents $\{K^{[np]}(i),\;
p\ge 1\}$ as
\[
K^{[np]}(i) =
a^{[np]}(i) + \sum_{k = 1}^pq_{k-1}^{(n,pn)}(i)z^{k(n-1)}a^{[(p-k)n]}(i).
\]
It is clear that $K^{[np]}(i)$ is of the form
\[
K^{[np]}(i) = z^{np} + {\cal O}(z^{(n-1)p-1}).
\]

\newtheorem{theoremdemo2}{Proposition}
\begin{theoremdemo2}
By virtue of (\ref{3})
\begin{equation}
z^{(1-n)p} K^{[np]}(i) = H^{(p)}(i).
\label{HK}
\end{equation}
\end{theoremdemo2}

\begin{pf}
First, notice that $z^{(1-n)p} K^{[np]}(i)$ and
$H^{(p)}(i)$ are Laurent series of the same form: $z^p + {\cal O}(z^{-1})$.
To proceed, it is enough to use the relation (\ref{3}) for $l=0$.
\end{pf}

Taking into consideration (\ref{HK}) we can write equations of DKP
chain restricted on submanifold $S_0^n$ as
\begin{equation}
z^{(n-1)p}\partial_p^{(n)}a(i) = a(i)
\left(K^{[np]}(i+1) - K^{[np]}(i)\right)
\label{edKP}
\end{equation}
with $\partial_p^{(n)}\equiv \partial/\partial t_p^{(n)}$.

Well known exactness property \cite{cfmp}
\[
\partial_qH^{(p)}(i) = \partial_pH^{(q)}(i)
\]
entails
\[
z^{(n-1)q}\partial_q^{(n)}K^{[np]}(i) = z^{(n-1)p}\partial_p^{(n)}K^{[nq]}(i).
\]
Therefore, one can introduce, for any integer $i$, the oscillatory function
$\Psi_i(z; t^{(1)}, t^{(2)}, ...) = \Psi_i(z, t)$ such that
\begin{equation}
H^{(p)}(i) = z^{(1-n)p}K^{[np]}(i) = \frac{\partial\ln\Psi_i}
{\partial t_p^{(n)}}.
\label{system}
\end{equation}

{\it Definition 1.} A Baker--Akhiezer function $\Psi_i(z, t)$ is any solution
of the system (\ref{system}) of the form
\[
\Psi_i(z, t) = z^{i}
\left(
1 + \sum_{k\ge 1}w_k(t)z^{-k}
\right)
\exp
\left(
\sum_{n\ge 1}\xi(t^{(n)}, z)
\right)
\]
with $\xi(t^{(n)}, z)\equiv \sum_{p\ge 1}t_p^{(n)}z^p$.

{\it Remark 1.} This function is defined up to the multiplication by a constant
monic Laurent series of degree $0$.

A bi-infinite sequence of $\Psi_i$'s forms the wavevector $\Psi = \{\Psi_i,\;
i\in{\bf Z}\}$. The relationship between the components of $\Psi$ and $a(i)$
and its discrete Fa\`a di Bruno iterates is provided by simple relation
\begin{equation}
a^{[k]}(i) = \frac{\Psi_{i+k}}{\Psi_i}.
\label{simple}
\end{equation}
Taking into consideration (\ref{relation}), for $r=1$, (\ref{system}) and (\ref{simple})
we arrive at the linear system
\begin{equation}
Q\Psi = z\Psi,\;\; z^{(n-1)p}\partial_p^{(n)}\Psi = Q_{+}^{pn}\Psi
\label{linear}
\end{equation}
with
\[
Q = \Lambda + \sum_{k\ge 1}q_{k-1}^{(n,1)}z^{k(n-1)}\Lambda^{r-kn}.
\]
The compatibility condition of the linear system (\ref{linear})
coincides with edKP Lax equation (\ref{lax-eq-m}).
So we can conclude that (\ref{edKP}) and (\ref{lax-eq-m}) are equivalent while
one-to-one correspondence $\{a_k(i)\}\leftrightarrow\{q_k^{(n)}(i)\equiv
q_k^{(n,1)}(i)\}$ is given by the relation
\[
z = a^{[1]}(i) + \sum_{k\ge 1}q_{k-1}^{(n)}(i)z^{k(n-1)}a^{[1-kn]}(i).
\]

\section{Reductions of edKP hierarchy}

As was observed in \cite{svinin1}, edKP hierarchy admits a class of
reductions specified by conditions $q_k^{(n,r)}(i)\equiv 0$ when $k\ge l$
with some $l\ge 1$. As a result, one can derive many well known integrable
discrete models like: generalized Toda lattices, Bogoyavlenskii lattices,
Belov--Chaltikian lattice and other known in the literature
(see \cite{svinin1}, \cite{svinin3} and references therein).
Let us show that these reductions can be interpreted
from geometric point of view simply as intersections of two invariant submanifolds
of DKP chain.

\begin{theoremdemo2}
Restriction of DKP chain on submanifold
$S_0^n\cap S_{l-1}^{ln-r}$ entails reduction of edKP hierarchy
specified by condition $q_k^{(n,r)}(i)\equiv 0$ for $k\ge l$.
\end{theoremdemo2}

\begin{pf}
The intersection $S_0^n\cap S_{l-1}^{ln-r}$ is defined
by two conditions
\begin{equation}
z^{1-n}a^{[n]}(i)\in H_{+}(i)\;\;\; {\rm and}\;\;\;
z^{l(1-n)+r}a^{[ln-r]}(i)\in H_{+}(i).
\label{two}
\end{equation}
Thus, we have to prove that (\ref{two}) leads to
\[
z^r = a^{[r]}(i) + \sum_{k=1}^lq_{k-1}^{(n,r)}(i)z^{k(n-1)}a^{[r-kn]}(i).
\]
The Laurent series $z^{l(1-n)+r}a^{[ln-r]}(i)$ is of degree $l$. From the
formula (\ref{3}) of the Lemma one derives that on $S_0^n\cap S_{l-1}^{ln-r}$
the latter can be uniquely represented as the linear combination\footnote{Since
on $S_0^n$ one has $H_{+}(i)={\rm span}(1, z^{1-n}a^{[n]}(i),
z^{2(1-n)}a^{[2n]}(i),...)$.}
\begin{equation}
z^{l(1-n)+r}a^{[ln-r]}(i) = z^{l(1-n)}a^{[ln]}(i) +
\sum_{k=1}^lc_{k-1}(i)z^{(l-k)(1-n)}a^{[(l-k)n]}(i)
\label{relation1}
\end{equation}
with some coefficients $c_k(i)$.
On the other hand shifting in (\ref{relation}) $i\rightarrow i+ln-r$
and multiplying then l.h.s. and r.h.s. by $z^{l(1-n)}a^{[ln-r]}(i)$ we have
\[
z^{r+l(1-n)}a^{[ln-r]}(i) = z^{l(1-n)}a^{[ln-r]}(i)a^{[r]}(i+ln-r)
\]
\[
+ \sum_{k\ge 1}q_{k-1}^{(n, r)}(i+ln-r)z^{(l-k)(1-n)}a^{[ln-r]}(i)
a^{[r-kn]}(i+ln-r).
\]
Taking into account (\ref{obvrel}) we obtain
\[
z^{r+l(1-n)}a^{[ln-r]}(i) = z^{l(1-n)}a^{[ln]}(i)
\]
\begin{equation}
+ \sum_{k\ge 1}q_{k-1}^{(n, r)}(i+ln-r)z^{(l-k)(1-n)}a^{[(l-k)n]}(i).
\label{ref}
\end{equation}
Comparing (\ref{relation1}) with (\ref{ref})
yields $c_k(i) = q_k^{(n, r)}(i+ln-r)$. Moreover, we see that the restrictions
(\ref{two}) yield $q_k^{(n,r)}(i)\equiv 0$, for $k\ge l$. Thus we complete
the proof.
\end{pf}

\section*{Acknowledgments}

I wish to thank Prof. Paolo Casati for bringing \cite{1,2} to my attention
and anonymous referee for carefully reading manuscript and for a number
of remarks which enabled the presentation of the paper to be improved.

\end{article}

\end{document}